\documentclass[reqno]{amsart}
\usepackage{amsmath,amssymb,latexsym}
\theoremstyle{plain}
\newtheorem{theorem}{Theorem}
\newtheorem{prop}{Proposition}

\theoremstyle{definition}

\theoremstyle{remark}

\begin{document}
\title{Locus configurations and $\vee$-systems}
%
%
     \author{O.A. Chalykh}
\address{Department of Mathematical Sciences,
        Loughborough University, Loughborough,
        Leicestershire, LE11 3TU, UK
        and Moscow State University, Moscow, Russia}
     \email{o.chalykh@lboro.ac.uk}
%
\thanks{The work has been supported by EPSRC (grant GR/M69548).}
%
       \author{A.P. Veselov}
\address{Department of Mathematical Sciences,
        Loughborough University, Loughborough,
        Leicestershire, LE11 3TU, UK
        and Landau Institute for Theoretical Physics, Moscow, Russia}
       \email{A.P.Veselov@lboro.ac.uk}
\begin{abstract}
We present a new family of the locus configurations which is not
related to $\vee$-systems thus giving the answer to one of the
questions raised in \cite{V1} about the relation between the
generalised quantum Calogero-Moser systems and special solutions
of the generalised WDVV equations. As a by-product we have new
examples of the hyperbolic equations satisfying the Huygens'
principle in the narrow Hadamard's sense. Another result is  new
multiparameter families of $\vee$-systems which gives new solutions of the
generalised WDVV equation.
\end{abstract}
\maketitle
\section{Introduction}
In the paper \cite{V1} (see also \cite{V2}) a mysterious relation
between the configurations
of hyperplanes appeared in the theory of generalised quantum
Calogero-Moser systems
(locus configurations \cite{CFV}) and the so-called $\vee$-systems
describing the special solutions of the generalised WDVV equations
has been observed. In this paper we investigate this relation further.

Our first result is a new family of the locus configurations, which
are not related to the WDVV equations (at least, in the way
described in \cite{V1}). This shows that the relation between the
locus configurations and $\vee$-systems is not general and is
applied only to a special subclass of the locus configurations,
thus answering one of the questions raised in \cite{V1}.

Another interesting feature of the new family is that it gives the
first examples
of the locus configurations in dimension more than two
for which corresponding Baker-Akhiezer functions do not satisfy the so-called
"old axiomatics" (see \cite{CFV} for details). It is plausible that
the subclass of the locus configurations related to $\vee$-systems is the one
with old axiomatics.

As a by-product we have new examples of hyperbolic equations
satisfying the Huygens' principle (in Hadamard's narrow sense).
The general relation between locus configurations and Huygens' principle
has been established in \cite{CFV}.
We should mention that as a particular two-dimensional case our family
contains the configuration first discovered by Yu.Berest and I.Lutsenko
in the relation with Huygens' principle \cite{BL}.

Another our result is two new multiparameter families of the $\vee$-systems in
dimension $n$. As a corollary we obtain new solutions of
the generalised WDVV equation.

\section{Generalised quantum CMS problem and locus configurations}

The famous Calogero-Moser-Sutherland integrable model describes a
pairwise interaction
of $N$ particles on the line with the potential which in
trigonometric case has the form
$$U = g\sum _{i<j}^{N}\frac{\omega^2}{\sin^2\omega(x_i - x_j)}.$$

Olshanetsky and Perelomov \cite{OP} were the first to consider the
generalisations
of this problem related to any root system $\mathcal R \in \bf{R}^n$.
Corresponding operator has the form
\begin{equation}
\label{1}
L = -\Delta + \sum\limits_{\alpha\in{\mathcal A}}
\frac{g_{\alpha}\omega^2}{\sin^2\omega(\alpha,x)}
\end{equation}
where $\mathcal A$ is a set of positive roots and $g_{\alpha}$ are
some parameters prescribed to
the roots in a way invariant under
the action of the corresponding Weyl group $W$.
Its quantum integrability in the sense of existence of $n$ pairwise commuting
quantum integrals has been shown in full generality first by Heckman and Opdam
(see \cite{H} and references therein).

In the papers \cite{CV},\cite{VSC} we have shown that if the
parameters have a special form $g_{\alpha} = m_\alpha(m_\alpha
+1)(\alpha,\alpha)$ with integer $m_\alpha$ then the operator
(\ref{1}) has more than $n$ quantum integrals, which by definition
means the algebraic integrability (see \cite{VSC} for precise
formulations). In that case the operator (\ref{1}) can be
intertwined with $L_0 = -\Delta$: there exists a differential
operator $\mathcal D$ such that $$L \mathcal D = \mathcal D L_0.$$

Surprisingly enough it turned out \cite{VFC} (see also \cite{CFV}) that the
same properties are valid for the operators (\ref{1}) for other
finite configurations $\mathcal A$. The first examples of such
non-root configurations have been found in \cite{VFC}.  They
consist of the following vectors in ${\bf R}^{n+1}$: $$
A_{n-1,1}(m)= \left\{
\begin{array}{lll}
e_i - e_j, &  1\le i<j\le n, & {\rm with \,\, multiplicity \,\,}
m^*,\\ e_i - \sqrt{m}e_{n+1}, &  i=1,\ldots ,n  & {\rm with \,\,
multiplicity \,\,}  1,
\end{array}
\right. $$ Here $m$ is an integer parameter, and the multiplicity
$m^*$ is the maximum of $m$ and $-1-m$. We should mention that in
\cite{CFV,VFC}
we have denoted this system as $A_n(m)$. The reason for the change of notation
will be clear from section 3 below.

The question of quantum algebraic integrability for the operators
\begin{equation}
\label{2}
L = -\Delta + \sum\limits_{\alpha\in{\mathcal A}}
\frac{m_\alpha(m_\alpha
+1)(\alpha,\alpha)\omega^2}{\sin^2\omega(\alpha,x)}
\end{equation}
and their rational limits
\begin{equation}
\label{3}
L = -\Delta + \sum\limits_{\alpha\in{\mathcal A}}
\frac{m_\alpha(m_\alpha +1)(\alpha,\alpha)}{(\alpha,x)^2}
\end{equation}
for a general finite set of noncollinear vectors $\mathcal A$ with
prescribed multiplicities led to the notion of the {\it locus
conditions} and {locus configurations} \cite{CFV}.

In the rational case we say that the potential
$$u(x) = \sum\limits_{\alpha\in{\mathcal A}} \frac{m_\alpha(m_\alpha
+1)(\alpha,\alpha)}{(\alpha,x)^2}$$
satisfies the locus conditions if
\begin{equation}
\label{loc}
u(x) - u(s_\alpha (x)) = O((\alpha,x)^{2m_\alpha})
\end{equation}
near every hyperplane $P_{\alpha}: (\alpha,x) = 0$. Here
$s_\alpha$ stands for the reflection with respect to this
hyperplane. In the trigonometric case we assume additionally that
the set $\mathcal A\subset \mathbb C^n$ generates a $\mathbb
Z$-lattice of rank $\le n$ and demand the locus conditions
(\ref{loc}) to be valid for any hyperplane $(\alpha,x) = \pi l, l
\in {\bf Z}$, where the potential has a pole. As it was shown in
\cite{C} these conditions are sufficient for the existence of the
intertwining operator $\mathcal D$ ( and thus for the algebraic
integrability).

We will call a finite set $\mathcal A$ of noncollinear vectors in
a (complex) Euclidean space a {\it locus configuration} if the
corresponding potential of the operator (\ref{2}) (and thus of the
operator (\ref{3})) satisfies the locus conditions (\ref{loc}).

The list of all known so far locus configurations in dimension
more than 2 is presented in \cite{CFV}. Besides the root systems
and the configuration $A_{n,1}(m)$ mentioned above it contains the
following family $$C_{n,1}(m,l) = \left\{
\begin{array}{lll}
e_i\pm e_j, &  1\le i<j\le n, &  {\rm with \,\, multiplicity \,\,}
k^*,\\ 2e_i, &  i=1,\ldots ,n  & {\rm with \,\, multiplicity \,\,}
m^*,\\ e_i\pm \sqrt{k}e_{n+1}, & i=1,\ldots ,n  &  {\rm with \,\,
multiplicity \,\,}  1,\\ 2\sqrt{k}e_{n+1} & {\rm with \,\,
multiplicity \,\,}  l^*,\\
\end{array}
\right. $$ Here $k,l,m$ are integer parameters related as $k =
\frac{2m+1}{2l+1}$, and $k^*, l^*, m^*$ have the same meaning as
in $A_{n,1}(m)$ case.

For all these configurations the corresponding Baker-Akhiezer
function \cite{CFV} satisfies the so-called "old axiomatics"
introduced in \cite{CV} which implies the new axiomatics
\cite{CFV} valid for any locus configuration. In dimension two
there are examples of the locus configurations first discovered by
Berest and Lutsenko \cite{BL} for which old axiomatics is not
valid. The question whether it is true or not in dimension more
than two was open until now. In the next section we will give the
answer to this question by presenting a new family of the locus
configurations which do not satisfy the old axiomatics. Actually
we will classify all the locus configurations of a certain type
and show that besides the known cases the list contains one new
interesting family.

\section{Locus configurations of $A$ type}

Let us consider a system $\mathcal A \subset \mathbb C^n$ which
consists of the vectors
\begin{equation}\label{a}
\alpha=\mu_ie_i-\mu_je_j\  (i<j)\qquad \text{with}\quad
m_\alpha=m_{ij}\in\mathbb Z_+\,.
\end{equation}
   Here $e_1,\dots, e_n$ is a
standard basis in $\mathbb C^n$ and $\mu_i\in\mathbb C^\times$ are
some prescribed parameters. We suppose that all the vectors
$\alpha\in\mathcal A$ are non-isotropic, i.e. $\mu_i^2+\mu_j^2\ne
0$ for all $i,j$. Altogether we have $n(n-1)/2+n$ parameters
$m_{ij},\mu_i$. The question we address here is when such a system
$\mathcal A$ is a locus configuration, i.e. when the potential in
\eqref{3} satisfies the locus conditions \eqref{loc}. Note that
the complex orthogonal group acts naturally on the set of all
locus configurations. In particular, the class of configurations
we consider is invariant under the action of the group
$W=S_n\ltimes (\mathbb Z_2)^n$ generated by permutations of the
coordinates and sign flips. In our classification below we will
not differ between configurations equivalent modulo $W$.

To start with, we recall that according to \cite{CFV} $\mathcal A$
is a locus configuration if and only if  all its two-dimensional
subsystems satisfy the locus conditions (see theorem 4.1 in
\cite{CFV}). So, let us start from considering a system $\mathcal
A^0$ consisting of three vectors $$\alpha=ae_1-be_2\,,\
\beta=be_2-ce_3\,,\ \gamma=ae_1-ce_3$$ with multiplicities
$(m_\alpha, m_\beta, m_\gamma)=(m,l,k)$.
\bigskip
\begin{prop}\label{n=3}
The system $\mathcal A^0\subset\mathbb C^3$ as above satisfies the
locus conditions in the following cases only:
\begin{align}\label{f1}
(1)&\quad a=b=c\quad\text{and}\quad m=l=k\qquad\text{(Coxeter
case)};\\\label{f2} (2)&\quad a=b\,,\ k=l=1\,,\ m>1\,, \
c=a\sqrt{m}\,;\\\label{f3} (3)&\quad a=b\,,\ k=l=1\,,\ m\ge 1\,,\
c=a\sqrt{-1-m}\,;\\\label{f4} (4)&\quad m=l=k=1\,,\
a^2+b^2+c^2=0\,.
\end{align}
(Notice that the partial case $m=1$ of $(3)$ appears also in the
family $(4)$.)
\end{prop}
\begin{proof}
This result is similar to the classification of all three-line
locus configurations on the plane (theorem 4.4 from \cite{CFV}),
and can be proven in a similar way. To begin with, notice that the
conditions \eqref{loc} are equivalent to the vanishing along
$\Pi_\alpha$ of some first odd normal derivatives of the potential
$u(x)$. More explicitly, they can be reformulated as follows: for
all $s=1,\dots, m_\alpha$
\begin{equation}\label{loc1}
\sum_{\beta\in\mathcal A\backslash
\alpha}\frac{m_\beta(m_\beta+1)(\beta,\beta)(\alpha,\beta)^{2s-1}}
{(\beta,x)^{2s+1}}\equiv 0 \qquad\text{along}\quad \Pi_\alpha \,.
\end{equation}
Applying this for our particular case, we arrive at the following
conditions:
\begin{equation}\label{loc2}
m_\beta(m_\beta+1)(\beta,\beta)(\alpha,\beta)^{2s-1}=
m_\gamma(m_\gamma+1)(\gamma,\gamma)(\alpha,\gamma)^{2s-1}\,\quad (
s=1,\dots, m_\alpha )\,,
\end{equation}
plus similar conditions for $\beta$ and $\gamma$. Suppose that
$m_\alpha>1$, then from the first two relations in \eqref{loc2} we
deduce immediately that $(\alpha,\beta)^2=(\alpha,\gamma)^2$, or
$a^2=b^2$, and hence $m_\beta=m_\gamma$. As a result, we see that
if at least two of the multiplicities
$(m_\alpha,m_\beta,m_\gamma)$ are greater than $1$, then they all
must be the same and $a^2=b^2=c^2$, which up to sign flips gives
us the Coxeter case $(1)$.

Another possibility is $(m_\alpha,m_\beta,m_\gamma)=(m,1,1)$ with
$m>1$. As we know already, in this case $a^2=b^2$ which provides
the conditions \eqref{loc2}. The two remaining locus conditions
are:
\begin{gather*}
m_\alpha(m_\alpha+1)(\alpha,\alpha)(\beta,\alpha)=
m_\gamma(m_\gamma+1)(\gamma,\gamma)(\beta,\gamma)\,,\\
m_\alpha(m_\alpha+1)(\alpha,\alpha)(\gamma,\alpha)=
m_\beta(m_\beta+1)(\beta,\beta)(\gamma,\beta)\,,
\end{gather*}
which gives $m(m+1)(a^2+b^2)b^2=2(c^2+a^2)c^2$ and leads to the
families $(2)$ and $(3)$.

Finally, for the remaining case
$(m_\alpha,m_\beta,m_\gamma)=(1,1,1)$ we have three locus
conditions
\begin{gather*}
(b^2+c^2)b^2=(c^2+a^2)a^2\\ (c^2+a^2)c^2=(a^2+b^2)b^2 \\
(a^2+b^2)a^2=(b^2+c^2)c^2\,.
\end{gather*}
It is easy to see that there are only two possibilities: either
$a^2=b^2=c^2$ or $a^2+b^2+c^2=0$. The first one gives us the
Coxeter case while the second one leads to the family $(4)$.
\end{proof}
\medskip
\medskip
\begin{theorem}\label{th1}
The system $\mathcal A\subset\mathbb C^n$ as in \eqref{a}
satisfies the locus conditions in the following cases only:
(1) $\mu_1=\dots =\mu_n$ and $m_{ij}\equiv m$ (Coxeter case);
(2) $\mu_1=\dots =\mu_{n-1}$, $m_{ij}\equiv m$ for all $i,j<n$,
$m_{in}=1$ for all $i$, $\mu_n= \mu_1\sqrt{m}$;
(3) $\mu_1=\dots =\mu_{n-1}$, $m_{ij}\equiv m$ for all $i,j<n$,
$m_{in}=1$ for all $i$, $\mu_n= \mu_1\sqrt{-1-m}$;
(4) $\mu_1=\dots =\mu_{n-2}=\mu$, $m_{ij}\equiv m$ for all $1\le
i,j\le n-2$, $m_{i,n-1}=m_{i,n}=m_{n-1,n}1$ for all $i\le n-2$,
$\mu_{n-1}= \mu\sqrt{m}$, $\mu_n=\mu\sqrt{-1-m}$;
(5) $n=3$, $m_{ij}\equiv 1$, $\mu_1^2+\mu_2^2+\mu_3^2=0$.
\end{theorem}
\bigskip
\begin{proof}
As we mentioned already, all we have to do is to check the locus
conditions for every two-dimensional subsystem in $\mathcal A$.
So, let us take any two-dimensional plane $\Pi$ in $\mathbb C^n$
and consider the set $\mathcal A_0=\mathcal A\cap \Pi$ assuming
that it is nonempty. Then there are three possibilities:

1) $\mathcal A_0$ consists of one vector;

2) $\mathcal A_0$ consists of two perpendicular vectors;

3) $\mathcal A_0$ consists of three vectors $\mu_ie_i-\mu_je_j,
\mu_je_j-\mu_ke_k, \mu_ie_i-\mu_ke_k$ for some $i<j<k$.

\noindent In the first two cases $\mathcal A_0$ is a locus system
for trivial reasons. So, essentially, the locus conditions for the
system $\mathcal A$ reduce to the requirement that for all $i<j<k$
the subsystem $\mathcal A_0$ as in 3) must be one of those listed
in Proposition 1.

Let us consider first the case $n=4$. We have four different
subsystems as in 3). The main point is that they cannot all be of
the type \eqref{f4}. Indeed, in that case we would have that
$$\mu_1^2+\mu_2^2+\mu_3^2=\dots=\mu_2^2+\mu_3^2+\mu_4^2=0$$ which
would imply that $\mu_1=\mu_2=\mu_3=\mu_4=0$. Thus, either all of
these subsystems are of the Coxeter type \eqref{f1}, or at least
one of them is of type \eqref{f2}--\eqref{f3}. In the latter case,
if we suppose that $\mu_2=\mu_3=\mu$ and $\mu_1=\sqrt{m}\mu$ then
for $\mu_4$ we have only two possibilities: either $\mu_4=\mu$, or
$\mu_4=\sqrt{-1-m}$. Therefore, all possible locus configurations
for $n=4$ are listed in the theorem. The general $n>4$ case
follows in a similar manner.
\end{proof}
\bigskip

Notice that we classified all the systems \eqref{a} for which the
corresponding {\it rational} potential \eqref{3} satisfies the
locus conditions. As a matter of fact, the {\it trigonometric}
version \eqref{2} of the resulting potentials will also satisfy
the locus conditions. This is enough to check for all
two-dimensional subsystems, which is not difficult. An explanation
for this phenomenon (that the weaker rational locus conditions
imply much stronger trigonometric ones) is due to the fact that
from the very beginning the system \eqref{a} was affine-equivalent
to the standard root system of type $A$.

As a result, the classification of all systems \eqref{a}
satisfying trigonometric locus conditions leads to the same list
from Theorem \ref{th1}.

The families $(2), (3)$ in Theorem \ref{th1} were found in
\cite{VFC,CFV}, it is the family $A_{n,1}(m)$ from the previous
section. The family $(4)$ is  new. We will denote as
$A_{n-1,2}(m)$ the corresponding system in $\mathbb C^{n+2}$:
$$
A_{n-1,2}(m)= \left\{
\begin{array}{lll}
e_i - e_j, &  1\le i<j\le n, & {\rm with \,\, multiplicity \,\,}
m\,,\\ e_i - \sqrt{m}e_{n+1}, &  i=1,\ldots ,n  & {\rm with \,\,
multiplicity \,\,}  1\,,\\ e_i - \sqrt{-1-m}e_{n+2}, &  i=1,\ldots
,n & {\rm with \,\, multiplicity \,\,}  1\,,\\
\sqrt{m}e_{n+1}-\sqrt{-1-m}e_{n+2} &  & {\rm with \,\,
multiplicity \,\,}  1\,.
\end{array}
\right. $$
Notice that for $m=1$ this system coincides with the
system $A_{n,1}(-2)$ from the previous section. Notice also that
$A_{n-1,2}(m)$ is contained in the hyperplane
$$x_1+\dots+x_n+\frac{1}{\sqrt
m}x_{n+1}+\frac{1}{\sqrt{-1-m}}x_{n+2}=0\,,$$ which leads to a
locus configuration in $\mathbb C^{n+1}$. In particular, for $n=2$
we obtain in this way a new locus configuration in dimension $3$.

\section{Generalised WDVV equations and $\vee$-systems.}

The generalised WDVV (Witten-Dijkgraaf-Verlinde-Verlinde) equations are
the following overdetermined system of nonlinear partial differential
equations:
\begin{equation}
F_iF_k^{-1}F_j=F_jF_k^{-1}F_i, \quad i,j,k=1,\ldots,n,
\label{wdvv}
\end{equation}
where $F_m$ is the $n\times n$ matrix constructed from the third partial
   derivatives of the unknown function $F=F(x^1,\ldots,x^n)$:
\begin{equation}
\label{f}
(F_m)_{pq}=\frac{\partial ^3 \, F}{\partial x^m \partial x^p \partial x^q}.
\end{equation}
In this form these equations have been presented by
A.Marshakov,A.Mironov and A.Morozov,
who showed that the Seiberg-Witten prepotential in $N=2$ four-dimensional
supersymmetric gauge theories satisfies this system \cite{MMM} (for more recent 
developments see \cite{Mir}).
Originally these
equations have appeared in topological field theory, their deep
geometry and relations
with integrable systems were first investigated by B.A.Dubrovin in \cite{D}.

In the papers \cite{V1},\cite{V2} the following special class of
solutions to (\ref{wdvv}):
\begin{equation}
F^{\mathfrak{A}}=\sum\limits_{\alpha \in \mathfrak{A}} (\alpha,x)^2
\, {\rm log} \,
(\alpha,x)^2,
\label{imF}
\end{equation}
where $\mathfrak{A}$ be a finite set of covectors $\alpha$ in the space $V^*$
dual to a vector space $V$, has been investigated.
It was shown that (\ref{imF}) satisfies the generalised WDVV
equations if $\mathfrak{A}$
satisfies certain conditions which led to the notion of $\vee$-systems.

Let us give the definition of the $\vee$-systems following \cite{V1}.

Let $V$ be a vector space (real or complex), $V^*$ be its dual,
$\mathfrak{A}\in V^*$
be a finite set of covectors which we assume to be non-collinear.
Let's introduce the following bilinear form
\begin{equation}
G^{\mathfrak{A}}=\sum\limits_{\alpha \in \mathfrak{A}} \alpha\otimes \alpha.
\label{mG}
\end{equation}
We will assume that the form $G^{\mathfrak{A}}$ is non-degenerate,
in the real case this equivalent to the fact that covectors $\alpha
\in \mathfrak{A}$
generate $V^*$.

This means that the natural linear mapping
$\varphi_\mathfrak{A} : V\rightarrow V^*$  defined by the formula
$$
(\varphi_\mathfrak{A}(u),v)=G^{\mathfrak{A}} (u,v), \, u,v \in V
$$
is invertible. We will denote  $\varphi_\mathfrak{A}^{-1}(\alpha),\,
\alpha \in V^*$ as
$\alpha^{\vee}$. By definition
$$\sum\limits_{\alpha \in \mathfrak{A}} \alpha^{\vee}\otimes \alpha = Id$$
as an operator in $V^*$ or equivalently
\begin{equation}
(\alpha,v)=\sum\limits_{\beta \in \mathfrak{A}}(\alpha, \beta^{\vee})(\beta,v).
\label{vee}
\end{equation}
for any $\alpha \in V^*, v \in V$.

Recall that for a pair of  bilinear forms $F$ and $G$ on the vector space $V$
one can define an eigenvector $e$ as the kernel of the bilinear form
$F-\lambda G$
for a proper $\lambda$:
$$
(F-\lambda G) (v,e)=0
$$
   for any $v \in V$. When $G$ is non-degenerate $e$ is the eigenvector
of the corresponding operator $\check F=G^{-1} F$: $$ \check
F(e)=G^{-1}F(e)=\lambda e. $$ Now let $\mathfrak{A}$ be as above
any finite set of non-collinear covectors $\alpha \in V^*$,
$G=G^\mathfrak{A}$ be the corresponding bilinear form (\ref{mG}),
which is assumed to be non-degenerate, $\alpha^\vee$ are defined
by (\ref{vee}). Define now for any two-dimensional plane $\Pi
\subset V^*$ a form
\begin{equation}
G^\mathfrak{A}_\Pi (x,y)= \sum\limits_{\alpha \in \Pi \cap
\mathfrak{A}} (\alpha,x)(\alpha,y).
\label{GP}
\end{equation}

{\bf Definition.} {\it We will say that $\mathfrak{A}$ satisfies the
$\vee$-conditions
if for any plane $\Pi \in V^*$ the vectors $\alpha^\vee, \, \alpha
\in \Pi \cap \mathfrak{A}$
are the eigenvectors of the pair of the forms $G^\mathfrak{A}$ and
$G^\mathfrak{A}_\Pi$.
In this case we will call $\mathfrak{A}$ as $\vee$-system.}
\smallskip

The $\vee$-conditions can be written explicitly as
\begin{equation}
\sum\limits_{\beta \in \Pi \cap \mathfrak{A}}
\beta(\alpha^\vee)\beta^\vee=\lambda \alpha^{\vee},
\label{expl}
\end{equation}
for any $\alpha \in \Pi \cap \mathfrak{A}$ and some $\lambda$, which may depend
on $\Pi$ and $\alpha$.

Geometrically we have three different cases:

1) If the plane $\Pi$ contains
no more than one covector from $\mathfrak{A}$ then $\vee$-conditions
are obviously satisfied (this means that these conditions should be
checked only for a finite
number of planes $\Pi$);

2) If the plane $\Pi$ contains only two covectors $\alpha$ and
$\beta$ from $\mathfrak{A}$ then the
condition (\ref{expl}) means that $\alpha^\vee$ and $\beta^\vee$ are
orthogonal with respect
to the form $G^\mathfrak{A}$:
$$
\beta(\alpha^\vee) = G^\mathfrak{A}( \alpha^\vee,\beta^\vee)=0;
$$

3) If the plane $\Pi$ contains more than two covectors from
$\mathfrak{A}$ this condition means that $G^\mathfrak{A}$ and
$G_\Pi^\mathfrak{A}$ restricted to the plane $\Pi^\vee \subset V$
are proportional:
\begin{equation}
\label{restr}
\left. G_\Pi^\mathfrak{A} \right|_{\Pi^\vee}=\lambda (\Pi) \left.
G^\mathfrak{A} \right|_{\Pi^\vee}.
\end{equation}

The $\vee$-conditions are known to be sufficient (in the real case -
necessary and sufficient)
for $F$ of the form (\ref{imF}) to satisfy the generalised WDVV (see
\cite{V1},\cite{V2}).

The natural examples of the $\vee$-systems are given by the
Coxeter systems consisting of the normals to the reflection
hyperplanes of some Coxeter group. It turned out that the locus
configurations can be used to construct non-Coxeter examples of
the $\vee$-systems. Namely for a given locus configuration
$\mathcal A$ consisting from the vectors $\alpha$ in the Euclidean
space $V$ with multiplicities $m_\alpha$ we can define a new set
of vectors in $V \approx V^*$ $$\mathfrak{A} = {\sqrt{m_\alpha}
\alpha, \alpha \in \mathcal A}.$$ A surprising fact discovered in
\cite{V1} is that for all locus configurations described in the
section 2 the corresponding sets $\mathfrak{A}$ satisfy the
$\vee$-conditions. A natural question arose whether this is a
common property of all locus configurations or not. Now we are
ready to answer this question.
\medskip
\begin{prop}
The systems $\mathfrak{A}$ corresponding to the new family of the
locus configurations $A_{n,2}(m)$ do not satisfy the
$\vee$-conditions. Corresponding function (\ref{imF}) is a
solution of the generalised WDVV equation only if $m=1$.
\end{prop}

To prove this let's notice that the definition of the $\vee$-systems
is affine invariant. This means that we can consider the projection
of the corresponding system $\mathfrak{A}$ into the hyperplane $x_{n+2} = 0,$
which has the form

$$
\mathfrak{A}_{n-1,2}(m)= \left\{
\begin{array}{lll}
\sqrt{m}(e_i - e_j), &  1\le i<j\le n,
\\ e_i - \sqrt{m}e_{n+1}, &  i=1,\ldots ,n
\\ e_i, &  i=1,\ldots ,n
\\ \sqrt{m}e_{n+1}.
\end{array}
\right. $$

A straightforward calculation shows that the $\vee$-conditions
corresponding to the plane containing the vectors
$e_i - \sqrt{m}e_{n+1}, e_i, \sqrt{m}e_{n+1}$ are satisfied if and
only if $m=1$
when we have the configuration of the type $A_{n,1}(-2)$.
Notice that since the system $\mathfrak{A}_{n-1,2}(m)$ is real
this implies that if $m \neq 1$ the corresponding function
(\ref{imF}) does not satisfy
the generalised WDVV equation according to the general result from \cite{V2}.

\medskip
In analogy with the previous section, it is natural to consider
$\vee$-systems of $A$ type which consist of the covectors
$\mu_{ij}(e_i-e_j)\in V^*$. Then the $\vee$-conditions imply some
algebraic relations on the parameters $\mu_{ij}$. In case $n=4$ we
are able to give the complete solution.

\begin{prop}
The system of $A_3$-type $$\mathfrak A=\{\mu_{ij}(e_i-e_j),\quad
1\le i<j\le 4\}$$ satisfies the $\vee$-conditions if and only if
$$\mu_{12}\mu_{34}=\mu_{13}\mu_{24}=\mu_{14}\mu_{23}\,.$$ The
corresponding family of solutions of the generalised WDVV equation
has the form
\begin{multline*}
F= c_1c_2(x_1-x_2)^2\log (x_1-x_2)^2 + c_2c_3 (x_2-x_3)^2\log
(x_2-x_3)^2 \\+ c_1c_3 (x_1-x_3)^2\log (x_1-x_3)^2+c_1 x_1^2\log
x_1^2 +c_2 x_2^2\log x_2^2+ c_3 x_3^2\log x_3^2\,,
\end{multline*}
with arbitrary $c_1,c_2,c_3$.
\end{prop}
It is interesting that this family of $\vee$-systems can be
extended to higher dimensions, though we are not sure whether or
not this exhausts all possibilities.

Namely, let us consider the system  $\mathfrak
{A}_n(c)=\{\sqrt{c_ic_j}(e_i-e_j),\quad 1\le i<j\le n+1\}$ in $\mathbb
R^{n+1}$ where $c_1,\dots, c_{n+1}$ are arbitrary (positive)
parameters. Without loss of generality, we may assume that
$c_{n+1}=1$ and restrict the system onto hyperplane $x_{n+1}=0$.
Thus we arrive at the following $n$-parametric family of
configurations in $\mathbb R^n$:
\begin{equation}\label{ch}
\mathfrak{A}_n(c)= \left\{
\begin{array}{lll}
\sqrt{c_ic_j}(e_i - e_j)\,, &  1\le i<j\le n\,,
\\ \sqrt{c_i}e_i\,, &  i=1,\ldots ,n\,.
\end{array}
\right.
\end{equation}

\bigskip
\begin{theorem} The system \eqref{ch} satisfies $\vee$-conditions
for any $c_1,\dots,c_n$. The corresponding family of solutions of
the generalised WDVV equation has the form
$$F=\sum_{i<j}c_ic_j(x_i-x_j)^2\log(x_i-x_j)^2+\sum_{i=1}^n
c_ix_i^2\log x_i^2\,.$$
\end{theorem}

\begin{proof}
Let us identify $V=\mathbb R^n$ with its dual using the standard
Euclidean structure. Then the bilinear form $G=G^{\mathfrak A}$
associated to the system \eqref{ch} according to the formula
\eqref{mG} looks as follows:
$$G(x,y)=\sum_{i<j}c_ic_j(x_i-x_j)(y_i-y_j)+\sum_{i=1}^n
c_ix_iy_i\,.$$ The associated matrix which we will denote by the
same symbol $G$ has the form $$G=(1+\sum_{i}c_i)C-c\otimes c\,,$$ where
$C={\rm diag}(c_1,\dots,c_n)$ and $(c\otimes c)_{ij}=c_ic_j$. A
straightforward check shows that its inverse has the form
$$G^{-1}=(1+\sum_{i}c_i)^{-1}(C^{-1}-e\otimes e)\,,$$ where $e=(1,\dots,1)$
and $(e\otimes e)_{ij}\equiv 1$ for all $i,j$.

To verify $\vee$-conditions, we should deal with two-dimensional
planes $\Pi$ containing at least two of the vectors
$\alpha,\beta\in \mathfrak A$. Altogether we have the following
$4$ different types of such planes:

(1) $\Pi=\langle e_i,e_j,e_i-e_j\rangle$;

(2) $\Pi=\langle e_i-e_j, e_j-e_k, e_i-e_k \rangle$;

(3) $\Pi=\langle e_i-e_j, e_k \rangle$;

(4) $\Pi=\langle e_i-e_j, e_k-e_l \rangle$.

Let us consider the first case. Let us fix a basis in $\Pi$ as
$\alpha=e_i$ and $\beta=e_j$. Then the corresponding plane
$\Pi^\vee$ is spanned by $\alpha^\vee=G^{-1}\alpha$ and
$\beta^\vee=G^{-1}\beta$. Using the explicit formula for $G^{-1}$
one easily finds that (up to a nonessential factor)
$$\alpha^\vee=(1,\dots,1+c_i^{-1},1,\dots,1)$$ (with $c_i^{-1}$
appearing in the $i$-th component) and similarly
$$\beta^\vee=(1,\dots,1+c_j^{-1},1,\dots,1)\,.$$ Now we should
check that the restrictions of the forms $G$ and $G_\Pi$ onto
$\Pi^\vee$ are proportional. Here $G_\Pi$ is given by the formula
$$G_\Pi(x,y)=c_jc_j(x_i-x_j)(y_i-y_j)+c_ix_iy_i+c_jx_jy_j\,.$$
After some calculations one finds that \begin{align*}
G(\alpha^\vee,\alpha^\vee)=&(1+\sum_k c_k)(1+c_i^{-1})\,,\\
G(\beta^\vee,\beta^\vee)=&(1+\sum_k c_k)(1+c_j^{-1})\,,\\
G(\alpha^\vee,\beta^\vee)=&1+\sum_k c_k\,.
\end{align*}
On the other hand, evaluating $G_\Pi$ we obtain that
\begin{align*}
G_\Pi(\alpha^\vee,\alpha^\vee)=&(1+c_i+c_j)(1+c_i^{-1})\,,\\
G_\Pi(\beta^\vee,\beta^\vee)=&(1+c_i+c_j)(1+c_j^{-1})\,,\\
G_\Pi(\alpha^\vee,\beta^\vee)=&1+c_i+c_j\,.
\end{align*}
This demonstrates that $G$ and $G_\Pi$ are proportional and gives
$\vee$-condition for the case (1).

In case (3) we take $\alpha=e_i-e_j$,\,$\beta=e_k$ and have only
to check that the Euclidean product $(\alpha^\vee,\beta)$ is zero.
The latter becomes obvious after calculating $\alpha^\vee$ which
is proportional to the vector $c_i^{-1}e_i-c_j^{-1}e_j$.

Two other cases are completely analogous. As a result, we conclude
that the system \eqref{ch} is a $\vee$-system for any values of
the parameters $c_1,\dots, c_n$ and the corresponding function
\eqref{imF} is a solution of the generalised WDVV equation.
\end{proof}

When $c_1=\dots=c_k$ for some $k<n$ and $c_{k+1}=\dots=c_n=1$ the
system \eqref{ch} reduces to the configuration $\mathcal
A_k*\mathcal A_l$ discovered by Berest and Yakimov (see
\cite{V2}). For general $c_i$ the constructed solutions of the
generalised WDVV equation seem to be new.

A natural question is what is the analogue of the family \eqref{ch}
for other classical root systems. The answer is given by the 
following family
\begin{equation}\label{bc}
\mathfrak{B}_n(c)= \left\{
\begin{array}{lll}
\sqrt{c_ic_j}(e_i \pm e_j)\,, &  1\le i<j\le n\,,
\\ \sqrt{2 c_i(c_i + c_0)}e_i\,, &  i=1,\ldots ,n\,.
\end{array}
\right.
\end{equation}
One can easily check that the $\vee$-conditions are satisfied for arbitrary
values of the parameters $c_0, c_1,\dots, c_n$.
The corresponding new solution of the generalised WDVV equation 
has the form
\begin{multline*}
F=\sum_{i<j}c_ic_j(x_i+x_j)^2\log(x_i+x_j)^2\\+
\sum_{i<j}c_ic_j(x_i-x_j)^2\log(x_i-x_j)^2+\sum_{i=1}^n
2c_i(c_i+c_0)x_i^2\log x_i^2\,.
\end{multline*}

\section{Relation to Huygens' Principle.}

Let us consider the second order hyperbolic equation
\begin{equation}\label{wave}
\mathcal L\phi(t,x)=0\,,\qquad \mathcal L=\Box_{N+1}+u(x)\,,
\end{equation}
where $\Box_{N+1}$ is the D'Alembert operator,
$\Box_{N+1}=\frac{\partial^2}{\partial
t^2}-\frac{\partial^2}{\partial x_1^2}-\dots
-\frac{\partial^2}{\partial x_N^2}$.

J.Hadamard raised the question when such an equation has the
fundamental solution located on the characteristic cone, or equivalently,
when it satisfyes the {\it Huygens' Principle} in the narrow Hadamard's sense.
For the review of the current situation with this problem
we refer to \cite{CFV}. In particular, the
theorem 6.1 from \cite{CFV} claims that if $u(x)$ is a real
rational potential \eqref{3} related to a locus configuration,
then the equation \eqref{wave} satisfies the Huygens' Principle
for large enough odd $N$. More
precisely, if $u(x)=u(x_1,\dots,x_n)$ is a potential \eqref{3}
related to a locus configuration $\mathcal A\subset \mathbb C^n$
then one should take $N\ge 2\sum_{\alpha\in\mathcal A}m_\alpha+3$.
Converse statement is also true if we assume
that all the Hadamard's coefficients are rational functions (see
theorem 6.2 in \cite{CFV}).

Applying this first for the configuration $A_{n-1,1}(m)$, we
arrive at the following potential $u$ depending on
$x_1,\dots,x_{n+1}$:
\begin{equation}\label{u1}
u=\sum_{i<j}^n 2m(m+1)(x_i-x_j)^{-2}+\sum_{i=1}^n 2(m+1)(x_i-\sqrt
m x_{n+1})^{-2}\,.
\end{equation}
For any $m\in\mathbb Z_+$ the corresponding equation \eqref{wave}
will satisfy the Huygens' Principle if $N$ is odd and $N\ge
mn(n-1)+2n+3$. When $m$ is negative integer, the potential
\eqref{u1} is no longer real-valued. However, one can make a
change of coordinates and think of $\sqrt{-1}x_{n+1}$ as a
$t$-variable. In this way we arrive at the time-dependent real
potential $u(t,x_1,\dots,x_n)$ as follows:
\begin{equation}\label{u2}
u=\sum_{i<j}^n 2m(m+1)(x_i-x_j)^{-2}+\sum_{i=1}^n 2(m+1)(x_i-\sqrt
{-m} t)^{-2}\,,
\end{equation}
and the corresponding huygensian equation \eqref{wave} for odd
$N\ge (-1-m)n(n-1)+2n+3$.

In case of the configuration $A_{n-1,2}(m)$ with $m\in\mathbb Z_+$
one can make a similar change of coordinates and think of
$\sqrt{-1}x_{n+2}$ as a $t$-variable. The corresponding potential
$u(t,x)$ will be of the form
\begin{multline}\label{uu}
u=\sum_{i<j}^n 2m(m+1)(x_i-x_j)^{-2}+\sum_{i=1}^n 2(m+1)(x_i-\sqrt
m x_{n+1})^{-2}\\- \sum_{i=1}^n 2m(x_i-\sqrt {m+1} t)^{-2}-2(\sqrt
m x_{n+1}-\sqrt{m+1}t)^{-2} \,.
\end{multline}
Thus, we arrive at the following result.

\begin{prop} The equation $(\Box_{N+1}+u(t,x))\phi=0$ with the
potential $u(t,x)$ given by \eqref{uu} with a positive integer $m$
satisfies the Huygens'
Principle for odd $N\ge mn(n-1)+4n+5$.
\end{prop}

This gives us the new examples of the huygensian equations, the first
of which appears in dimension $N=17$ for $n=2$ and $m=2$.

\end{document}